\documentclass[prl,twocolumn,aps,showpacs,superscriptaddress]{revtex4}
\usepackage{amsmath}
\usepackage{amsfonts}
\usepackage{amssymb}
\usepackage{yfonts}
\usepackage{graphicx}
\usepackage{dcolumn}  
\begin{document}
\title{Discrete and continuum spectra in the unified shell model approach}
\author{Alexander Volya}
\affiliation{Department of Physics,
Florida State University, Tallahassee, FL 32306-4350, USA}
\author{Vladimir Zelevinsky}
\affiliation{NSCL and Department of Physics and Astronomy,
Michigan State University, East Lansing, MI 48824-1321, USA}
\date{\today}
\begin{abstract}
A new version of the nuclear shell model unifies the consideration of the
discrete spectrum, where the results agree with the standard shell
model, and continuum. The ingredients of the method are the
non-Hermitian effective Hamiltonian, energy-dependent one-body and
two-body decay amplitudes, and self-consistent treatment of
thresholds. The results for helium and oxygen isotope chains well
reproduce the data.
\end{abstract}
\maketitle

The standard nuclear shell model (SM) approach erects a wall between the
description of intrinsic structure and reactions since it does not
account for the continuum spectrum. This problem became acute due
to experimental progress towards nuclei far from stability. The
proximity of continuum in loosely bound nuclei influences all
their properties which makes necessary a common description of the
discrete and continuum spectrum. The classical book
\cite{mahaux69} formulated the unified approach to nuclear
reactions based on the SM. However, only recently realistic
practical methods \cite{bennaceur99,betan02,michel02,michel03}
were suggested. Below we formulate a version of the shell model in
continuum (CSM) that naturally includes the conventional SM with
discrete spectrum and gives in the same framework the description
of reaction cross sections and decay channels with no restriction
by single-nucleon decays. We show non-trivial applications to the
chains of helium and oxygen isotopes. The formalism is based on
the Feshbach method \cite{feshbach58} of projecting out the states
$|c;E\rangle$ of ``external" $Q$-space, corresponding to the decay
channels $c$ at continuum energy $E$, and constructing the
effective Hamiltonian ${\cal H}$ that acts in the ``internal"
$P$-space of the many-body SM states $|1\rangle$.

Within the total space $P+Q$ we solve the Schr\"odinger equation
$H|\alpha\rangle=E|\alpha\rangle  $, where the full wave function
$|\alpha\rangle$ is a superposition of internal states $|1\rangle$
and external states $|c;E\rangle$. The elimination of external
states leads to the closed equation for the internal part with the
effective Hamiltonian
\begin{equation} {\cal H}(E)=H+\Delta (E) -\frac{i}{2}\,W (E)\,,
                                               \label{3}
\end{equation}
where the principal value term $\Delta$ is due to the off-shell
processes of virtual excitation into channel space $Q$,
\begin{equation} \langle 1 |\Delta (E)|2\rangle = \sum_{c} \,{\cal
P} \int dE' \, \frac{{A^c_1(E')}^*\, {A^c_2(E')}}{E-E'},
                                                  \label{4}
\end{equation}
and the explicitly non-Hermitian term
\begin{equation} \langle 1 | W(E)|2\rangle =
2\pi  \sum_{c\,({\rm open})} {A^c_1(E)}^*\,{A^c_2 (E)} \label{5}
\end{equation}
represents on-shell decays into the channels open at given energy.
The amplitudes $A_{1}^{c}(E)$ are the matrix elements $\langle
c;E|H|1\rangle$ of the original Hamiltonian between $Q$ and $P$
spaces. The two new terms depend on running energy $E$ so that we
deal with a non-Hermitian and energy-dependent Hamiltonian. It is
important that the observable quantities, such as the scattering
matrix, can be written in terms of the same amplitudes and the
Hamiltonian ${\cal H}$ \cite{PRC67}.
The factorized form (\ref{5}) of $W$ follows from the
unitarity of the $S$-matrix.

This formulation is exact. The complex poles of the analytical
continuation of $S(E)$, ${\cal
E}_{\alpha}=E_{\alpha}-(i/2)\Gamma_{\alpha}$, are resonances.
Since the decay amplitudes $A_{1}^{c}(E)$ must vanish below
threshold energy $E_{c}^{{\rm th}}$, their energy behavior
complicates the situation making decay non-exponential. On the
other hand, this ensures a continuous matching to the conventional
SM: below thresholds the same bound states are obtained assuming
that $\Delta(E)$ is included into renormalization of the SM
interaction. In this work we use $m$-scheme Slater determinants on
bound single-particle orbitals in the mean field potential for
$P$-space basis. The channel states are defined by their
asymptotic quantum numbers. A one-particle channel of an $N$-body
system with quantum numbers $j$, energy of the particle
$\epsilon_{j}$, and the residual nucleus in the state
$|\alpha;N-1\rangle$, so that $E=E_{\alpha}+\epsilon_{j}$, is
labelled as
$|c\rangle=b^{\dagger}_{j}(\epsilon_{j})|\alpha;N-1\rangle$. The
total spin-isospin quantum numbers are restored by the solution if
the Hamiltonian respects this symmetry.

We assume that one-body decays are determined by the
single-particle part $h$ of the full SM Hamiltonian,
\begin{equation} A^c_1(E_\alpha+\epsilon_j)=
\sum_\nu a^j_\nu(\epsilon_j)\,\langle
\alpha;N-1|b_\nu|1;N\rangle,                       \label{7}
\end{equation}
where $a^j_{\nu}(\epsilon)$ describes the single-particle
transition from the SM state $\nu$ into continuum state $j$ with
energy $\epsilon_{j}$, mediated by the mean-field interaction $h.$
If the valence space is small, and each single-particle state is
uniquely marked by spin-isospin quantum numbers, only one orbital
$\nu$ can couple to a given continuum channel, $a^j_\nu(\epsilon)=
\delta_{\nu j} a^j(\epsilon)$; then $\nu$ can be identified with a
continuum index $j$. The non-Hermitian part of effective
Hamiltonian becomes ($E=\epsilon_j+E_\alpha$)
\begin{equation}
\langle 1 | W(E)|2\rangle = 2\pi \delta_{1 2} \sum_{c\,({\rm
open})} |a^{j}(\epsilon_j)|^2 |\langle
\alpha;N-1|b_j|1;N\rangle|^2.                   \label{9}
\end{equation}

Being in general a many-body operator, in some important cases
$W$ is effectively reduced to a single-particle form.
Thus, far from thresholds one can ignore the energy dependence
and use the closure to simplify the summation in (\ref{9}),
\begin{equation} \langle 1 | W(E)|2\rangle = 2\pi \delta_{1 2}
\sum_{j} |a^{j}|^2\,
|\langle 1;N|b^\dagger_j b_j|1;N\rangle|^2.       \label{10}
\end{equation}
As a result, $W$ becomes a one-body operator that assigns a width
$\gamma_j=2 \pi \left |a^{j}\right |^2$ to each unstable
single-particle state $j=\nu$ and can be combined with the SM
Hamiltonian by introducing complex single-particle energies
$e_\nu=\epsilon_\nu-i \gamma_\nu/2$. A similar picture emerges if
the residual two-body interaction is weak, and the sum over
daughter systems $\alpha$ in Eq. (\ref{9}) is dominated by a
single term. The single-particle interpretation of the continuum
coupling $W$ is generally valid when the removal of a particle
does not lead to a significant restructuring of the mean field. In
the lowest order \cite{PRC67}, the phenomenon of decay does not
change the structure of the internal wave function. The width of
the many-body state $\alpha$ is then given by the expectation
value $\Gamma_\alpha=\langle \alpha |W| \alpha\rangle$; if $W$ is
assumed to be a one-body operator, $\Gamma_\alpha=\gamma_\nu
(\epsilon_\nu) \langle \alpha |b^\dagger_\nu b_\nu| \alpha
\rangle.$ This defines a many-body decay width as a product of a
single-particle width and the spectroscopic factor $\langle \alpha
|b^\dagger_\nu b_\nu| \alpha \rangle$ .

The {\sl single-particle} decay follows from the one-body
scattering problem in an average potential $V$ that enters the
one-body Hamiltonian $h$; this determines the reduced
single-particle amplitudes $a^j_\nu(\epsilon)$. Assuming spherical
symmetry of $V(r)$, the single-particle decay amplitude is found
as $a^l(\epsilon)=\int_0^\infty \, F_l(r) V(r) u_l(r)\, dr,$ where
$u_{l}$ is the radial function, and (for a neutral particle)
$F_{l}(r)=(2\mu/\pi k)^{1/2}kr\,j_{l}(kr)$. For the near-threshold
cases, the continuum admixtures are dominated by the long
wavelength states; with the $k\rightarrow 0$ behavior $F_l(r)\sim
(kr)^{l+1}$, then $a^l(\epsilon) \sim \,\epsilon^{(l+1)/2}$.

The {\sl two-nucleon decay} admixes the two-body continuum in the
asymptotic form $|c\rangle = b_j^\dagger(\epsilon)
b^\dagger_{j'}(\epsilon') |\alpha;N-2 \rangle$. The channel state
is characterized by energies of emitted particles, $\epsilon$
and $\epsilon'$. With the one-body interaction used for the
single-particle decay, the admixture from the two-particle
continuum appears as a second order contribution,
\begin{equation}
A^c_1(E)= \sum_\beta a^j(\epsilon) a^{j'}(\epsilon') \left
(\frac{(b_{j})_{\alpha\beta}(b_{j'})_{\beta 1}}
{E-E_\beta-\epsilon}+ (j\leftrightarrow j')\right )\,, \label{21}
\end{equation}
that proceeds through intermediate states $\beta$ of a nucleus
with $N-1$ particles. Eq. (\ref{21}) and corresponding
contribution to $W$, Eq. (\ref{5}), allow for additional
simplifications in the near-threshold region. The overall width
behaves as $\gamma\sim q^{2+l+l'}$, where $q$ is the total
available kinetic energy.

In contrast to this {\sl sequential} decay amplitude, a {\sl
direct} two-body transition requires the presence of a two-body
interaction in the Hamiltonian. To describe this process, a pair
amplitude is introduced, $A^c_1(E)=a^{(L)}(\epsilon_1,
\epsilon_2)\,\, \langle \alpha;N-2|p_L|1;N \rangle\, $ where
operator $p_L=\{b_\nu \otimes b_{\nu'}\}_L$ removes a pair, and
only the quantum numbers $L$ of the pair are conserved. In the
long wavelength limit, the two-body decay amplitude scales with
energy in accordance with the phase space volume, identically to
the sequential decay. The dominant contribution of orbital
momentum $L=0$ results in  $W \sim q^2$; this ``pairing'' channel
is also favored by the short-range nature of residual forces.

\begin{figure}
\begin{center}
\includegraphics[width=3.3 in]{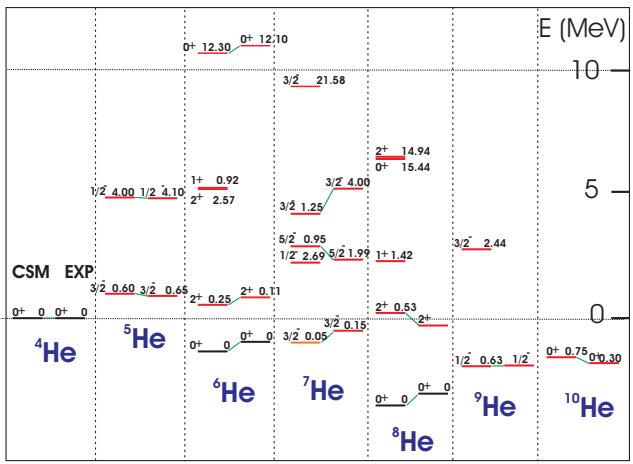}
\vspace{-0.2 cm}
\end{center}
\caption{Results for He isotopes. For each isotope CSM result is
on the left, while experimental value is to the right;
experimentally observed states are linked to theoretical
counterparts. The decay width is shown on the right in the units
of MeV. Energies (or centroids of decaying states) are not shown,
but the energy scale is given on the vertical axis.\label{he2}}
\vspace{-0.2 cm}
\end{figure}

As a first application, we consider the chain of helium isotopes
from $^4$He to $^{10}$He. The internal $P$-space contains two
single-particle levels, $p_{3/2}$ and $p_{1/2}$. The interaction
and single-particle energies are defined in
\cite{cohen65,stevenson88}. For the one-body channels, using the
Woods-Saxon potential with the parameters adjusted for $^5$He, it
was determined that, even for several MeV above threshold, the
single-particle amplitudes can be described by the
parameterization $\gamma_{3/2}(\epsilon)=0.608
\,\epsilon^{3/2}\,$MeV and $\gamma_{1/2}(\epsilon)=0.3652\,
\epsilon^{3/2}\,$MeV for the decay width from $p_{3/2}$ and
$p_{1/2}$ states, respectively. The sequential two-body decay is
computed using Eq. (\ref{21}) with the near-threshold
approximation for the energy dependence of single-particle
amplitudes. The direct two-body decay is introduced only for the
pair emission with total angular momentum $L=0$. It is assumed
that all internal $L=0$ pairs couple to the continuum with the
same amplitude $a^{(L=0)}(\epsilon_1, \epsilon_2).$ The direct
two-body amplitude is parameterized as $a^{(L=0)}(\epsilon_1,
\epsilon_2)= (\epsilon_1+\epsilon_2)/{3\sqrt{2\pi}}$, where the
numerical constant fixes the strength of the residual two-body
interaction. This is the only parameter of the model, and it was
adjusted to two-body decays of $^6$He. Figure \ref{he2} and Table
\ref{hetab} show the results of the CSM calculation in a good
agreement with data.
\begin{table}
{\tiny
\begin{tabular}{|c|c||c|c|c|c|c||c|c|c|c|}
\hline
A&   J &     E(SM) &    E(a)&  E (b)&  E(CSM)&    E(EX)&  $\Gamma$(a) &  $\Gamma$(b)& $\Gamma$(CSM)&    $\Gamma$(EX)\\
\hline
4&  0&  0&  0&  0&  0&  0&  0&  0&  0&  0\\
5&  3/2&    0.992&  0.992&  0.992&  0.992&  0.895& 0.6&    0.6&    0.6&    0.648\\
5&  1/2&    4.932&  4.932&  4.932&  4.932&  4.895& 4&  4&  4&  4.1\\
6&  0&  -1.379& -1.379& -1.379& -1.379& -0.973&    0&  0&  0&  0\\
6&  2&  0.515&  0.515&  0.529&  0.529&  0.825& 0&  0.248&  0.248&  0.113\\
6&  2&  4.745&  5.25&   5.25&   5.25&   &   2.566&  2.566&  2.566&  \\
6&  1&  5.889&  5.32&   5.32&   5.32&   &   0.922&  0.922&  0.922&  \\
6&  0&  11.088& 10.911& 10.911& 10.803& 11.128& 5.532&  5.532&  12.303& 12.1\\
7&  3/2&    -1.016& -1.016& -1.016& -1.016& -0.528&    0.046&  0.046&  0.046&  0.15\\
7&  1/2&    2.24&   2.239&  2.253&  2.253&  &   2.357&  2.689&  2.69&   \\
7&  5/2&    2.85&   2.888&  2.911&  2.911&  2.393& 0.727&  0.944&  0.944&  1.99\\
7&  3/2&    4.495&  4.379&  4.222&  4.22&   5.273& 0.541&  1.113&  1.246&  4\\
7&  3/2&    10.223& 8.857&  9.521&  9.544&  &   7.818&  16.379& 21.578& \\
8&  0&  -3.591& -3.591& -3.591& -3.591& -3.108&    0&  0&  0&  0\\
8&  2&  0.19&   0.196&  0.191&  0.19&   -0.308&    0.231&  0.506&  0.53&   \\
8&  1&  2.427&  2.304&  2.331&  2.321&  &   1.026&  1.455&  1.418&  \\
8&  0&  6.376&  6.003&  6.527&  6.489&  &   5.286&  3.456&  15.449& \\
8&  2&  6.882&  6.839&  6.538&  6.572&  &   2.283&  13.86&  14.94&  \\
9&  1/2&    -1.992& -1.992& -1.992& -1.992& -1.958&    0.634&  0.634&  0.634&  \\
9&  3/2&    2.805&  2.801&  2.802&  2.797&  &   1.557&  2.425&  2.443&  \\
10& 0&  -1.649& -1.649& -1.649& -1.649& &   0.073&  0.504&  0.746&  \\
\hline
\end{tabular}
} \caption{\label{hetab}Comparison of the traditional SM and CSM
with data for He isotopes. First two columns identify the mass
number and spin of the state. Next five columns compare energies:
$E$(SM) traditional SM; $E$(a) version of CSM with only one-body
decays included; $E$(b) version of CSM with one-body decay and its
second order contribution to the two-body process; $E$(CSM) full
CSM including the direct two-body decay mode; $E$(EX) experimental
data, if available. Next five columns compare decay widths from
CSM calculations with experimental values; the SM calculation
gives only discrete energies. All numbers are given in MeV;
energies are measured from the ground state of $^4$He.
 }
\vspace{-0.2 cm}
\end{table}

Several features are worth emphasizing. (i) By design of the model
with $\Delta(E)$ approximated by a constant and included in the
adjusted SM Hamiltonian, energies of bound states agree exactly
with the results of the standard SM. The ground states of $^4$He,
$^6$He and $^8$He are nucleon-stable in agreement with experiment.
(ii) Energies of resonances deviate from the SM predictions $-$
the continuum is restructuring internal states. For narrow
separated resonances the effect is small. However, in systems
strongly coupled to continuum, internal phase transitions with
formation of broad (super-radiant) and very narrow states can be
found, see \cite{pentaquark} and references therein. A similar
effect can be traced in Table \ref{hetab} so that decaying states
are ``pushed'' further into continuum. (iii) A detailed comparison
reveals information about structure and dominant decay modes. The
decay of the $2^+$ state in $^6$He is a sequential two-body
process leading to the ground state of $^4$He with no effects of
direct $L=2$ pair emission. This is not the case for the 0$^+_{2}$
state, where one-body and sequential two-body decays reproduce
only about half of the observed width. The rest comes from the
direct two-body emission of $L=0$ neutron pair to the ground state
of $^4$He favored since the $0^+_{2}$ state is mainly a coherent
excitation of a neutron pair to the $p_{1/2}$ orbital. The largest
deviation from experiment is seen for the $3/2^-$ state in $^7$He
(although the data also have a large uncertainty). Possibly, the
direct pair decay to $3/2^-$ state in $^5$He (as seen from Table,
it is very important) is followed by a fast further breakdown to
$^4$He. Sequential three-body decays were not considered in the
model. The admixture of the high-lying $3/2^-$ state with a large
width is also possible. (iv) The results agree with calculations
\cite{michel02,michel03} by a different method but with a similar
mean field and schematic residual force.

\begin{figure*}
\begin{center}
\includegraphics[width=7 in]{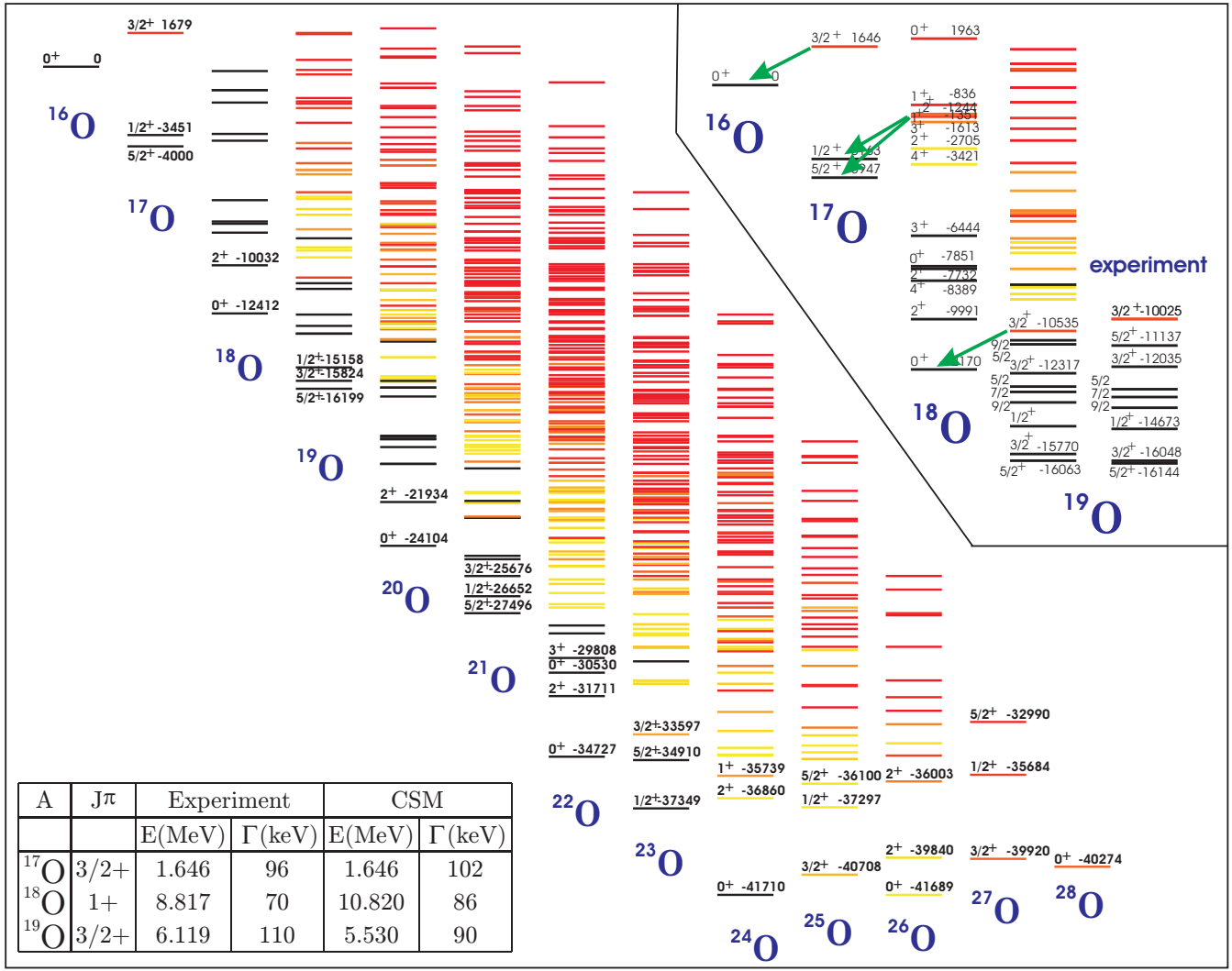}
\vspace{-0.2 cm}
\end{center}
\caption{CSM calculations for oxygen isotopes with the HBUSD
interaction. States from yellow (long lifetime) to red (short
lifetime) are resonance states. The insert on the upper right
shows a more detailed picture for the lightest $^{16}$O to
$^{19}$O isotopes. Decays from all states that are experimentally
measured are shown with arrows. A full comparison between
available data and the calculation is given for $^{19}$O. Energies
are expressed in units of keV. Comparison of widths with available
data is given in the table in lower-left corner. For both inserts
the interaction USD was used that works better in this mass
region. \label{os2} } \vspace{-0.2 cm}
\end{figure*}

As second application, Fig. \ref{os2}, we perform a full
calculation (all states and all interaction matrix elements in the
$sd$-SM included) for the chain of oxygen isotopes; the first
results were reported earlier \cite{PRC67}. As for He isotopes,
self-consistency is established at all stages: decay energies for
given parent-daughter pairs are consistent with the reaction
processes; decay amplitudes for given states are in agreement with
the reaction calculation and SM spectroscopic factors; the effect
of decay on intrinsic states is accounted for in the
diagonalization of the non-Hermitian energy-dependent Hamiltonian.
The model is essentially parameter-free. The standard SM
interaction (USD \cite{USD} or its slightly modified version for
heavier isotopes HBUSD \cite{HBUSD}) was used supplemented by a
parameterization of one-body continuum coupling found from
scattering off the Woods-Saxon potential adjusted for $^{17}$O.
Due to scarce experimental data on neutron decays in heavy
isotopes, only sequential two-body processes are considered which
requires no additional parameters. A comparison with data is shown
in the inserts on Fig. \ref{os2} for the USD interaction. A number
of features similar to those discussed for the helium isotopes can
be noticed.

We presented fully realistic CSM calculations and compared results
with experiment. The main ingredients of the method are the SM
with a good effective interaction in the discrete spectrum, the
non-Hermitian effective Hamiltonian for open channels, correct
energy dependence of the decay amplitudes, and the self-consistent
calculation of thresholds for a chain of isotopes. The high
quality SM results for discrete states are reproduced by the CSM
exactly. The comparison with experiment indicates similar high
quality for resonances in the continuum. A model requires very few
parameters, and most of them can be fixed via simplified
calculations of scattering processes. The self-consistency between
reactions and structure through energy dependence, thresholds, and
parent-daughter relations is a part of the method. Consideration
of continuum goes beyond single-particle processes. A detailed
study of one- and two-body decays and comparison with data provide
insight into structure of resonant states and interplay of
intrinsic and continuum dynamics.

The authors acknowledge support from the U. S. Department of
Energy, grant DE-FG02-92ER40750;
Florida State University FYAP
award for 2004, and National Science Foundation, grants
PHY-0070911 and PHY-0244453. Help and advices from B. Alex Brown
are highly appreciated.
\vspace{-0.2 cm}

\end{document}